\documentclass[conference]{IEEEtran}
\IEEEoverridecommandlockouts

\usepackage{cite}
\usepackage{amsmath,amssymb,amsfonts}
\usepackage{algorithmic}
\usepackage{graphicx}
\usepackage{textcomp}
\usepackage{xcolor}
\usepackage{booktabs}
\usepackage{array}
\usepackage{booktabs}
\usepackage{hyperref}

\def\BibTeX{{\rm B\kern-.05em{\sc i\kern-.025em b}\kern-.08em
    T\kern-.1667em\lower.7ex\hbox{E}\kern-.125emX}}
\begin{document}

\title{Federated Learning on the American Science Cloud using APPFL
}

\author{
\IEEEauthorblockN{
Zilinghan Li\IEEEauthorrefmark{1},
Abhijit Chunduru\IEEEauthorrefmark{1}\IEEEauthorrefmark{2},
Harinarayan Krishnan\IEEEauthorrefmark{3}, 
Eric Chagnon\IEEEauthorrefmark{3},\\
Peter Nugent\IEEEauthorrefmark{3}, 
Kibaek Kim\IEEEauthorrefmark{1}, 
Ravi Madduri\IEEEauthorrefmark{1}
}
\IEEEauthorblockA{
\IEEEauthorrefmark{1}Argonne National Laboratory
\IEEEauthorrefmark{2}University of Massachusetts Amherst
\IEEEauthorrefmark{3}Lawrence Berkeley National Laboratory
}
\IEEEauthorblockA{
\{zilinghan.li, schunduru, kimk, madduri\}@anl.gov, \{hkrishnan, echagnon, penugent\}@lbl.gov
}
}

\maketitle

\begin{abstract}
The American Science Cloud (AmSC), established under the Genesis Mission of the U.S. Department of Energy (DOE), aims to integrate DOE high-performance computing systems, experimental facilities, and data resources into a single, coordinated, AI-driven discovery platform. AmSC's early services focus on curated artifacts, such as gated inference access to hosted models, experiment tracking, and function execution across computing facilities. However, what these services lack is a means to train a model across organizational boundaries where data cannot be centralized due to policy, privacy, or scale. This is, by definition,  a use case for federated learning (FL) and a growing class of scientific AI. In this paper, we show that this gap can be bridged by deploying the orchestration logic of the Advanced Privacy-Preserving Federated Learning (APPFL) framework as a scalable cloud service on top of the primitives AmSC already provides: project-scoped authentication that supports secure and reliable federation membership, function execution that drives distributed training at each site, experiment tracking that records round-level performance, and finally, the model-hosting and inference infrastructure that can be leveraged to distribute the federated trained models to authorized participants. We argue that offering federated computing as an important AmSC service would unlock privacy-constrained scientific collaborations, enabling public-private partnerships in model building  while exercising and enhancing the platform's own federated infrastructure. 
\end{abstract}

\begin{IEEEkeywords}
Federated Learning, Multi-Facility Computing, American Science Cloud, High-Performance Computing, Privacy-Preserving Computing
\end{IEEEkeywords}

\section{Introduction}

The U.S. Department of Energy (DOE) has launched the Genesis Mission, a national initiative to connect its computing, data, and experimental facilities into a unified AI-driven discovery system. The enabling platform, the American Science Cloud (AmSC), 
is a federated, cloud-like service layer over existing DOE high-performance computing (HPC) facilities and the Energy Sciences Network (ESnet). AmSC's early services emphasize hosting and serving, for example, a model access gateway offering inference over hosted commercial and open-weight models, model cataloging and experiment tracking, an interface for cross-facility function execution and agentic job management, and shared data-catalog services~\cite{amscdocshome2026}.

This host-and-serve model, while valuable, presumes that either the data or the trained model can be generally centralized. However, the collaborations AmSC is ultimately meant to serve, i.e., cross-institutional scientific AI spanning national laboratories, universities, and industry partners, frequently violate such presumption: protected information, export-controlled measurements, and proprietary industrial data can be neither pooled nor shipped to a central trainer. Federated learning (FL)~\cite{mcmahan2017communication,kairouz2021advances} is the workflow that remains once such centralization is off the table: rather than gathering data into one place, FL sends the model to the data, computes local updates in place, and exchanges and aggregates only model parameters or gradients. It is a multi-facility scientific workflow that AmSC's current services are not primarily geared toward. Our position is that {federated computing} would be an important addition to AmSC's services, and, more consequentially, that AmSC can provide it with little new infrastructure development effort. 

The capabilities such a service needs, including secure identity management, distributed execution, experiment tracking, and model distribution, are already present in AmSC's deployed services in different forms. The missing piece is a thin orchestration layer that composes them into an iterative federated training loop, which can be provided by APPFL, the Advanced Privacy-Preserving Federated Learning framework developed at Argonne National Laboratory~\cite{ryu2022appfl,li2025advances}. APPFL was designed for the deployment regime that AmSC targets: FL across HPC resources, institutional clusters, and commercial cloud environments, using communication back-ends such as gRPC and Globus Compute~\cite{chard2020funcx} that traverse institutional boundaries, with authentication handled through federated identity. 

The primary contributions of this paper are threefold. First, we identify a service gap in AmSC's current catalog: it can serve, invoke, and track curated artifacts, but lacks means to support the iterative, privacy-preserving, cross-institutional training that a large class of scientific AI requires. Second, we present a component-level design that bridges this gap by mapping each part of a production FL framework (APPFL) onto primitives AmSC already provides, thus showing that federated computing can be realized largely by composition rather than by building new infrastructure. Third, we derive from this mapping a small set of capabilities that AmSC needs to extend to better support servicing federated training.

\section{Background and Related Work}

\subsection{APPFL: Cross-Silo Federated Learning for Science}
APPFL is an open-source FL framework designed for cross-silo scientific deployments. Its architecture is organized into four components with distinct responsibilities in federated training loop. (1) The \emph{server} implements aggregation strategies ranging from synchronous FedAvg~\cite{mcmahan2017communication} to asynchronous and heterogeneity-aware algorithms such as FedCompass~\cite{li2024fedcompass} and FedQueue~\cite{li2026fedqueue}, which accounts for disparities in client computing capability and resource queuing time, common conditions when clients are HPC allocations at different facilities. (2) The \emph{client} handles local training and is independent of the server orchestration logic. (3) The \emph{communicator} abstracts the transport layer over multiple backends, including gRPC for request-response cross-site data exchange and Globus Compute~\cite{chard2020funcx} for federated function execution across administrative domains, with authentication provided by Globus-based federated identity~\cite{tuecke2016globus}. (4) The \emph{privacy} layer applies differential-privacy mechanisms and secure aggregation protocols to the exchanged model updates to further protect from information leakage. In addition, APPFL also provides a hosted service layer (APPFLx) through which institutions join federations via web-based authentication~\cite{hoang2025enabling}, reducing the operational burden of cross-institutional studies in domains such as biomedicine to reduce the entry barrier to launching FL experiments for domain experts.

\subsection{Related Work}
There exist several other open-source frameworks that target cross-silo FL, including FedML~\cite{he2020fedml}, Flower~\cite{beutel2020flower}, NVIDIA FLARE~\cite{roth2022nvidia}, and OpenFL~\cite{reina2021openfl}, alongside APPFL. Similar to APPFL, each framework is a largely self-contained stack: to run a federation, it supplies not only the learning logic but also its own identity handling, cross-site transport, and job orchestration. Our work is orthogonal to this line. Rather than proposing another framework, we ask how an existing science platform such as AmSC can provide these surrounding capabilities once through a service, so that any such framework reduces to the FL logic that is genuinely its own, and the resulting composition benefits the ecosystem broadly rather than a single framework. We adopt APPFL as the concrete reference because it aligns most closely with the AmSC setting. It targets cross-silo scientific federations rather than cross-device deployments, matching AmSC's model of institutions and HPC centers as clients. It also supports heterogeneity-aware and queue-aware schedulers, FedCompass and FedQueue, directly address the disparate compute capacities and batch-queue wait times that dominate round-completion time across DOE facilities, and its reliance on Globus Compute and Globus-based federated identity already exercises the cross-domain execution and authentication that AmSC must provide, making the position of this paper concrete rather than hypothetical.

This perspective connects FL to the broader study of scientific workflows on multi-facility infrastructure. The workflows community has long examined the orchestration, fault-tolerance, and data-management demands of execution that spans administrative domains~\cite{da2021community,da2024workflows}, and work applying the FAIR principles to HPC-resident AI artifacts~\cite{wilkinson2025applying} anticipates the provenance requirements that iterative federated training places on a model store. Within DOE, the Integrated Research Infrastructure (IRI) program~\cite{miller2023integrated}, on which AmSC's research services build~\cite{doe2026scac}, identified the cross-facility patterns that federated training instantiates in a particularly demanding form. To our knowledge, this is the first work to propose FL as a composition of AmSC's platform services.

\section{Federated Computing as a Composition of AmSC Primitives}
\label{sec:mapping}
\begin{figure*}[!t]
\centering
\includegraphics[width=0.8\textwidth]{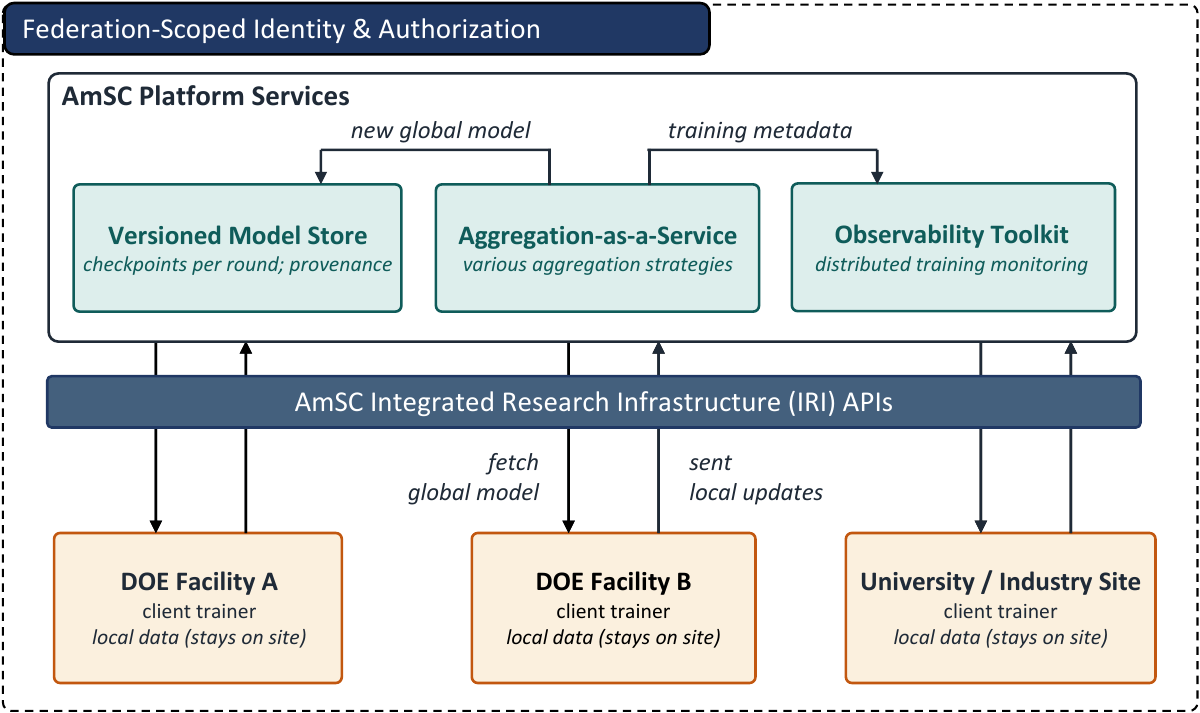}
\caption{Federated training assembled from AmSC services. Client trainers at DOE facilities and external university or industry sites fetch the current global model and send local updates through the AmSC Integrated Research Infrastructure (IRI) APIs, under federation-scoped identity and authorization; raw data never leave the participating sites. On the platform side, aggregation-as-a-service produces each new global model and writes it to a versioned model store that keeps per-round checkpoints and provenance, while an observability toolkit collects training metadata across the federation. Framework such as APPFL supplies the FL logic that drives this loop.}

\label{fig:architecture}
\end{figure*}

An FL framework supplies two very different kinds of functionality: (1) the federation logic such as aggregation strategies, client scheduling, and privacy mechanisms, and (2) the generic distributed systems infrastructure that surrounds it, such as identity management, data transfer, training execution, and experiment monitoring. For APPFL, like every cross-silo framework, provides both itself. Our design aims to keep the first with APPFL and integrates the second through the AmSC ecosystem: each of APPFL's components is mapped to the AmSC service that can host it, so that a federation is assembled from AmSC primitives rather than run as a self-contained stack. Table~\ref{tab:mapping} summarizes this correspondence, distinguishing capabilities AmSC already provides and those it partially provides. Fig.~\ref{fig:architecture} shows the resulting workflow. 

\begin{table*}[htbp]
\caption{Mapping FL System Components onto AmSC Services}
\label{tab:mapping}
\centering
\renewcommand{\arraystretch}{1.15}
\begin{tabular}{
@{}
    >{\raggedright\arraybackslash}p{3.4cm}
    >{\raggedright\arraybackslash}p{4.5cm}
    >{\raggedright\arraybackslash}p{8.2cm}@{}}
\toprule
\textbf{FL component} &
\textbf{Closest AmSC capability} &
\textbf{Status and required extension} \\
\midrule
Aggregation server &
Inference services &
\emph{Prototype.} Requires on-demand launching of aggregation services. \\
Client execution &
HPC allocations via AmSC APIs &
\emph{Partial.} Reliable job scheduling is supported on some DOE leadership computing facilities. \\
Federated identity &
AmSC identity provider; project-scoped tokens &
\emph{Fulfilled.} AmSC identity provider or tokens can be used to authorize FL experiments. \\
Privacy layer &
Secure enclaves &
\emph{Addition.} Enclaves provides additional protection to data at rest. \\
Monitoring and observability &
Facility-local monitoring &
\emph{Partial.} MLFlow provides partial observability, but lacks monitoring tools to oversee the whole distributed workflows. \\
Versioned global hosting &
Model access gateway &
\emph{Partial.} Inference-oriented; requires versioned, writable artifacts \\
\bottomrule
\end{tabular}
\end{table*}

\subsection{From Inference-as-a-Service to Aggregation-as-a-Service}
AmSC's inference services host models on DOE resources in response to authenticated requests. Federated aggregation or other FL-related operation is likewise a function invoked on demand when client updates arrives.  To prototype this, we have already developed an on-demand federated aggregation service on NERSC's Spin~\cite{nersc_spin}, which allows authorized users to launch an aggregation server on-demand through a Jupyter notebook for FL clients to connect to run FL experiments (\href{https://portal.trac.appflx.link/}{https://portal.trac.appflx.link/}). We propose to generalize it into \emph{aggregation-as-a-service}: a platform interface in which aggregation strategies are executed on updates sent by connected authorized clients. Frameworks such as APPFL would contribute strategies as plugins instead of operating standalone servers, so that collaborations without a facility to host an aggregator could still launch FL experiments as needed.

\subsection{Client Execution and Scheduling}
FL clients at HPC facilities run today as batch jobs or Globus Compute endpoints that the framework must keep available across rounds. AmSC already exposes reliable job scheduling on some DOE leadership computing facilities to some authorized users, which covers part of this need, but two problems remain. First, this support is not uniform across the diverse facilities a federation may span. Second, batch schedulers are designed for few long-running jobs, however, FL training usually consists of many short, recurring ones for timely global synchronization, so round-completion time is dominated by queue delays and straggling clients rather than by computation or transfer. A recurring-reservation or event-triggered execution primitive in AmSC's API layer would let a federation specify that local training at a facility runs whenever a new global model version appears, which is a platform-level contract in place of per-site workarounds.

\subsection{Federated Identity}
Identity, the prerequisite for any cross-institutional data exchange, is the capability that AmSC fully provides. Its project-scoped tokens, issued through the platform identity provider~\cite{amscdocshome2026}, already suffice to authorize the participants of a federated experiment: a token scoped to a project can gate who may join a federation and deposit or retrieve its models. Existing APPFL and APPFLx deployments authenticate through Globus federated identity~\cite{hoang2025enabling}, so a bridge between the two regimes would let current federations adopt platform identity incrementally, an integration convenience rather than a blocking gap. What remains on the communication side is a writable transport for the model updates themselves, which spares each deployment from separately negotiating firewall traversal and authentication.

\subsection{Privacy Mechanisms and Secure Enclaves}
On the one hand, the FL framework already provides algorithmic privacy layers, including differential privacy applied to model updates, and secure aggregation, which prevents the aggregator from inspecting individual contributions. On the other hand, the AmSC's secure enclaves can serve as additional privacy enhancements: executing aggregation inside an enclave adds a hardware trust anchor when federation members do not fully trust the hosting facility, complementing the framework's cryptographic protections.

\subsection{Cross-Facility Observability}
Cross-facility federated training is a distributed workflow for which no complete view of the execution state currently exists. Each facility monitors only the jobs executing on its own resources, while the FL framework captures only the events recorded in its own logs. Consequently, the global state of a training run, including round progression, per-client staleness, communication timings, and failure attribution, cannot be reconstructed by any single party. AmSC's experiment tracking (MLflow) captures per-run metrics and so provides partial visibility, but it does not reconstruct this federation-wide state.Our development and operation of HiveWatch (\href{https://github.com/APPFL/hivewatch}{https://github.com/APPFL/hivewatch}), an observability toolkit for general distributed training that can be easily integrated into FL framework such as APPFL, has shown that the availability of such cross-site telemetry is a determining factor in whether failures within a federation can be diagnosed and attributed. What is needed at the platform level is a standardized telemetry format for federated training rounds that covers round events, client participation, update provenance, and communication timings, all integrated with facility-level monitoring so that operators and scientists share a single view of a federation's state.

\subsection{Model Hosting as the Federation's Shared State}
After an FL training experiment, the server maintains the authoritative global model between training rounds. On AmSC, this state maps to a versioned platform artifact: the global model is checkpointed at each round, and provenance records link each version to the participating clients and the aggregation metadata. Current AmSC model services are not yet oriented toward this pattern: the Model Access Gateway provides authenticated access to hosted models for inference, but no user-writable artifact store. What federated training needs is read/write access to versioned artifacts with append-only lineages, whose round-level provenance also aids reproducibility. Realizing this raises several questions: which artifact granularity, such as full models, deltas, or compressed updates, is appropriate for repeated exchange over the wide-area network; what provenance schema adequately captures a training lineage; and how responsibility for such a store should be divided between AmSC model services and the High Performance Data Facility, whose early-access system is being designed around AmSC data services~\cite{doe2026scac}.

\section{Conclusion}
The American Science Cloud is being built now, and the services it settles on will shape a decade of scientific AI. We have argued that federated computing belongs in that catalog, and that it does not need to be built from scratch: identity, function execution, experiment tracking, and model hosting already exist as AmSC primitives, and a production framework, APPFL, supplies the FL logic that composes them into an iterative training loop. Mapping APPFL onto these primitives shows how much is already in place and isolates the capabilities that remain to be extended: aggregation-as-a-service; recurring, event-triggered client execution; cross-facility observability; and writable, federation-scoped model exchange. Closing these gaps would let AmSC serve the privacy-constrained collaborations that a host-and-serve catalog cannot, while turning FL from a fragile, self-hosted stack into a first-class platform service. We offer the mapping and agenda here as a starting point.

\section*{Acknowledgment}
This material is based upon work supported by the US
Department of Energy, Office of Science, under contract number DE-AC02-06CH11357.

\bibliographystyle{IEEEtran}
\bibliography{references}

\end{document}